\documentclass{article}
\usepackage{cite}
\usepackage{graphicx}
\usepackage{dcolumn}

\begin{document}

\date{}
\title{On the permutation symmetry of atomic and molecular wavefunctions}
\author{Francisco M. Fern\'{a}ndez \thanks{%
E-mail: fernande@quimica.unlp.edu.ar} \\
INIFTA, Divisi\'on Qu\'imica Te\'orica\\
Blvd. 113 S/N, Sucursal 4, Casilla de Correo 16, \\
1900 La Plata, Argentina}
\maketitle

\begin{abstract}
In this paper we analyze a recently proposed approach for the
construction of antisymmetric functions for atomic and molecular
systems. It is based on the assumption that the main problems with
Hartree-Fock wavefunctions stem from their lack of proper
permutation symmetry. This alternative building approach is based
on products of a space times a spin function with opposite
permutation symmetry. The main argument for devising such factors
is that the eigenfunctions of the non-relativistic Hamiltonian are
either symmetric or antisymmetric with respect to the
transposition of the variables of a pair of electrons. However,
since the eigenfunctions of the non-relativistic Hamiltonian are
basis for the irreducible representations of the symmetric group
they are not necessarily symmetric or antisymmetric, except in the
trivial case of two electrons. We carry out a simple and
straightforward general analysis of the symmetry of the
eigenfunctions of the non-relativistic Hamiltonian and illustrate
our conclusions by means of two exactly-solvable models of $N=2$
and $N=3$ identical interacting particles.
\end{abstract}

\section{Introduction}

\label{sec:intro}

It is well known that the application of quantum mechanics to systems of
identical particles requires taking into account the permutation symmetry of
the Hamiltonian operator\cite{CTDL77}. The postulates of quantum mechanics
state that the wavefunctions that describe Bosons or Fermions should be
symmetric or antisymmetric, respectively, with respect to the transposition
of the variables of identical particles\cite{CTDL77}. For this reason, the
solutions to the Schr\"{o}dinger equation for atomic and molecular systems
are commonly approximated by Slater determinants that are antisymmetric
functions constructed from monoelectronic spin-orbitals\cite{P68}. This way
of constructing the variational functions leads to the Hartree-Fock (HF)
method and its variants\cite{P68}. Besides, it is well known that the Slater
determinants are consistent with Pauli exclusion principle\cite{P68}.

In a recent paper Nascimento\cite{N19} argues that the main problems
exhibited by the HF approximate wavefunctions stem from the fact that they
do not actually take into account the proper permutation symmetry of the
non-relativistic atomic and molecular Hamiltonians. He proposes that
alternative better functions should be the product of a space function times
a spin one with opposite permutation symmetry, namely, symmetric$\times $%
antisymmetric or antisymmetric$\times $symmetric. According to the
author this more suitable approach is based on the fact that each
exact
eigenfunction $\psi _{el}$ of the non-relativistic electronic Hamiltonian $%
H_{el}$ satisfies $P_{ij}\psi _{el}=\pm \psi _{el}$ $\forall i,j$, where $%
P_{ij}$ is a permutation (transposition\cite{CTDL77}) operator for a pair of
particle labels $i$ and $j$. The author argues that the better results
derived from the Generalized Valence Bond (GVB) Method with respect to HF
comes precisely from the fact that the former takes into account true
permutation symmetry while the latter does not.

The purpose of this paper is to analyze the validity of the argument based
on the permutation symmetry of $\psi _{el}$ just mentioned. In section~\ref
{sec:Perm sym} we outline the main ideas about the application of
permutation groups to atomic and molecular systems. In section~\ref
{sec:solvable models} we illustrate the general results by means of two
exactly-solvable models for $N=2$ and $N=3$ identical interacting particles.
Finally, In section~\ref{sec:conclusions} we summarize the main results, add
some further comments and draw conclusions.

\section{Permutation symmetry}

\label{sec:Perm sym}

The non-relativistic Hamiltonian $H$ of an $N$-electron system is invariant
under the transposition $P_{ij}$ of any pair of electron variables. For this
reason the commutator between both operators vanishes: $[H,P_{ij}]=0$. Since
any permutation of the electron labels can be written as a product of a
finite number of transpositions\cite{CTDL77} we focus on the latter.
However, we must keep in mind that the symmetric group $S_{N}$ consists of $%
N!$ permutation operators of which $N(N-1)/2$ are transpositions\cite{CTDL77}%
.

If $\psi $ is an eigenfunction of $H$ with eigenvalue $E$ then the obvious
equalities $P_{ij}H\psi =HP_{ij}\psi =EP_{ij}\psi $ tell us that $P_{ij}\psi
$ is also an eigenfunction of $H$ with eigenvalue $E$. If the energy level $%
E $ is non-degenerate then $P_{ij}\psi =\lambda \psi $, where $\lambda $ is
a constant. Since $P_{ij}^{2}$ is the identity operator then $\lambda =\pm 1$
and $\psi $ is either symmetric or antisymmetric under the permutation of
the pair $i,j$ of electron coordinates. However, if the energy level is
degenerate, then $P_{ij}\psi $ and $\psi $ are not necessarily linearly
dependent and we cannot assure that $P_{ij}\psi =\pm \psi $. The question is
if it is possible to find linear combinations of the degenerate
eigenfunctions that are either symmetric or antisymmetric. In what follows
we will show that the answer is negative for most of them.

In the case of two electrons we can find a set of eigenfunctions common to $%
H $ and $P_{12}$ and, consequently, we are certain that all the
eigenfunctions of $H$ can be chosen to be either symmetric or antisymmetric.
However, when $N>2$ the transposition operators do not commute $%
[P_{ij},P_{kl}]\neq 0$\cite{CTDL77} and we cannot obtain eigenfunctions
common to $H$ and all the transposition operators. In the next section we
will show an exactly-solvable example that illustrates these points. In the
case of three particles the equations
\begin{eqnarray}
P_{12}P_{13}f(\mathbf{r}_{1},\mathbf{r}_{2},\mathbf{r}_{3}) &=&P_{12}f(%
\mathbf{r}_{3},\mathbf{r}_{2},\mathbf{r}_{1})=f(\mathbf{r}_{3},\mathbf{r}%
_{1},\mathbf{r}_{2}),  \nonumber \\
P_{13}P_{12}f(\mathbf{r}_{1},\mathbf{r}_{2},\mathbf{r}_{3}) &=&P_{13}f(%
\mathbf{r}_{2},\mathbf{r}_{1},\mathbf{r}_{3})=f(\mathbf{r}_{2},\mathbf{r}%
_{3},\mathbf{r}_{1}),
\end{eqnarray}
clearly show that $P_{12}$ and $P_{13}$ do not commute. If we choose $\psi$
to be eigenfunction of both $H$ and $P_{12}$ it will not be necessarily
eigenfunction of $P_{13}$ or $P_{23}$.

The symmetric group $S_{3}$ for three identical particles is isomorphic to
the point groups $D_{3}$ and $C_{3v}$ commonly used in the description of
the symmetry of molecular states in quantum-chemistry calculations based on
the Born-Oppenheimer approximation\cite{P68}. The three irreducible
representations are labelled $A_{1}$, $A_{2}$ and $E$\cite{C90} (the
construction of the character table for the symmetric group $S_{3}$ can be
seen in\newline
https://groupprops.subwiki.org/wiki/Determination\_of\_character\_table\_of%
\_symmetric\_group:S3).\newline
We can use the character table of $C_{3v}$ to obtain the basis functions for
the irreducible representations by straightforward application of the
projection operators $P_{S}$\cite{C90} to an arbitrary function $f(\mathbf{r}%
_{1},\mathbf{r}_{2},\mathbf{r}_{3})$. For example:
\begin{eqnarray}
f_{A_{1}}(\mathbf{r}_{1},\mathbf{r}_{2},\mathbf{r}_{3}) &=&P_{A_{1}}f(%
\mathbf{r}_{1},\mathbf{r}_{2},\mathbf{r}_{3})=\frac{1}{6}\left[ f\left(
\mathbf{r}_{1},\mathbf{r}_{2},\mathbf{r}_{3}\right) +f\left( \mathbf{r}_{3},%
\mathbf{r}_{2},\mathbf{r}_{1}\right) +f\left( \mathbf{r}_{1},\mathbf{r}_{3},%
\mathbf{r}_{2}\right) \right.   \nonumber \\
&&\left. +f\left( \mathbf{r}_{2},\mathbf{r}_{1},\mathbf{r}_{3}\right)
+f\left( \mathbf{r}_{3},\mathbf{r}_{1},\mathbf{r}_{2}\right) +f\left(
\mathbf{r}_{2},\mathbf{r}_{3},\mathbf{r}_{1}\right) \right] ,  \nonumber \\
f_{A_{2}}(\mathbf{r}_{1},\mathbf{r}_{2},\mathbf{r}_{3}) &=&P_{A_{2}}f(%
\mathbf{r}_{1},\mathbf{r}_{2},\mathbf{r}_{3})=\frac{1}{6}\left[ f\left(
\mathbf{r}_{1},\mathbf{r}_{2},\mathbf{r}_{3}\right) +f\left( \mathbf{r}_{3},%
\mathbf{r}_{2},\mathbf{r}_{1}\right) +f\left( \mathbf{r}_{1},\mathbf{r}_{3},%
\mathbf{r}_{2}\right) \right.   \nonumber \\
&&\left. -f\left( \mathbf{r}_{2},\mathbf{r}_{1},\mathbf{r}_{3}\right)
-f\left( \mathbf{r}_{3},\mathbf{r}_{1},\mathbf{r}_{2}\right) -f\left(
\mathbf{r}_{2},\mathbf{r}_{3},\mathbf{r}_{1}\right) \right] ,  \nonumber \\
f_{E_{1}}(\mathbf{r}_{1},\mathbf{r}_{2},\mathbf{r}_{3}) &=&P_{E}f(\mathbf{r}%
_{1},\mathbf{r}_{2},\mathbf{r}_{3})=\frac{1}{3}\left[ 2f\left( \mathbf{r}%
_{1},\mathbf{r}_{2},\mathbf{r}_{3}\right) -f\left( \mathbf{r}_{2},\mathbf{r}%
_{3},\mathbf{r}_{1}\right) -f\left( \mathbf{r}_{3},\mathbf{r}_{1},\mathbf{r}%
_{2}\right) \right] ,  \nonumber \\
f_{E_{2}}(\mathbf{r}_{1},\mathbf{r}_{2},\mathbf{r}_{3}) &=&P_{E}f(\mathbf{r}%
_{2},\mathbf{r}_{1},\mathbf{r}_{3})=\frac{1}{3}\left[ 2f\left( \mathbf{r}%
_{2},\mathbf{r}_{1},\mathbf{r}_{3}\right) -f\left( \mathbf{r}_{1},\mathbf{r}%
_{3},\mathbf{r}_{2}\right) -f\left( \mathbf{r}_{3},\mathbf{r}_{2},\mathbf{r}%
_{1}\right) \right] ,  \nonumber \\
f_{E_{3}}(\mathbf{r}_{1},\mathbf{r}_{2},\mathbf{r}_{3}) &=&P_{E}f(\mathbf{r}%
_{3},\mathbf{r}_{1},\mathbf{r}_{2})=\frac{1}{3}\left[ 2f\left( \mathbf{r}%
_{3},\mathbf{r}_{1},\mathbf{r}_{2}\right) -f\left( \mathbf{r}_{1},\mathbf{r}%
_{2},\mathbf{r}_{3}\right) -f\left( \mathbf{r}_{2},\mathbf{r}_{3},\mathbf{r}%
_{1}\right) \right] ,  \nonumber \\
f_{E_{4}}(\mathbf{r}_{1},\mathbf{r}_{2},\mathbf{r}_{3}) &=&P_{E}f(\mathbf{r}%
_{1},\mathbf{r}_{3},\mathbf{r}_{2})=\frac{1}{3}\left[ 2f{\left( \mathbf{r}%
_{1},\mathbf{r}_{3},\mathbf{r}_{2}\right) }-f{\left( \mathbf{r}_{2},\mathbf{r%
}_{1},\mathbf{r}_{3}\right) }-f{\left( \mathbf{r}_{3},\mathbf{r}_{2},\mathbf{%
r}_{1}\right) }\right] .  \nonumber \\
&&
\end{eqnarray}
We clearly see that from the six functions derived from all the permutations
of the variables of $f(\mathbf{r}_{1},\mathbf{r}_{2},\mathbf{r}_{3})$ we
obtain one of symmetry $A_{1}$, one of symmetry $A_{2}$ and two pairs of
symmetry $E$.\ One can easily verify that $P_{ij}f_{A_{1}}=f_{A_{1}}$, $%
P_{ij}f_{A_{2}}=-f_{A_{2}}$ and $P_{ij}f_{E_{i}}\neq \pm f_{E_{i}}$. We
cannot derive symmetric or antisymmetric functions from linear combinations
of $f_{E_{1}}$, $f_{E_{2}}$, $f_{E_{3}}$ and $f_{E_{4}}$ because $%
P_{A_{1}}f_{E_{i}}=P_{A_{2}}f_{E_{i}}=0$. The conclusion is that the
eigenfunctions of $H$ having symmetry $E$ will not be eigenfunctions of all
the transposition operators which contradicts Nascimento's assumption that $%
P_{ij}\psi =\pm \psi $ for any eigenfunction $\psi $ of $H$ \cite{N19}. We
will illustrate these points in section~\ref{sec:solvable models}. It should
be taken into account that the six symmetry-adapted functions just discussed
are nonzero provided that the functions obtained by the six permutations of
the three arguments of $f(\mathbf{r}_{1},\mathbf{r}_{2},\mathbf{r}_{3})$ are
linearly independent.

The situation is even worse for systems of more particles; for example, $%
S_{4}$ is isomorphic to $O$ or $T_{d}$\cite{C90} (see also\newline
https://groupprops.subwiki.org/wiki/Linear\_representation\_theory\_of%
\_symmetric\_group:S4)\newline
and we have the irreducible representations $A_{1}$, $A_{2}$, $E$, $T_{1}$
and $T_{2}$, so that only the functions that are basis for $A_{1}$ and $A_{2}
$ are eigenfunctions of all the transposition operators. In this case we
cannot obtain symmetric or antisymmetric functions from linear combinations
of the basis functions for the remaining irreducible representations $E$, $%
T_{1}$ and $T_{2}$.

The \textit{full} molecular Hamiltonian (which includes the kinetic energy
of nuclei) is invariant under permutation of the variables of the electrons
and also of identical nuclei. Both the atomic and full molecular
Hamiltonians are also invariant under parity $Pf(\mathbf{r}_{1},\mathbf{r}%
_{2},\ldots )=f(-\mathbf{r}_{1},-\mathbf{r}_{2},\ldots )$. Since $[H,P]=0$
and $[P,P_{ij}]=0$ then all the eigenstates of $H$ satisfy $P\psi =\pm \psi $%
; that is to say, they are either even or odd.

\section{Exactly-solvable models}

\label{sec:solvable models}

In the case of atoms and molecules one cannot solve the Schr\"{o}dinger
equation exactly and, consequently, one is restricted to apply the results
outlined above to the approximate wavefunctions that one commonly uses to
describe the physical properties of such systems. However, there are many
simple toy models that are exactly solvable and exhibit the desired
permutation symmetry. In what follows, we illustrate the main ideas of the
preceding section by means of two such examples.

\subsection{$N=2$ toy model}

The simple two-particle model
\begin{equation}
H=-\frac{1}{2}\left( \frac{\partial ^{2}}{x_{1}^{2}}+\frac{\partial ^{2}}{%
x_{2}^{2}}\right) +\frac{1}{2}\left( x_{1}^{2}+x_{2}^{2}\right) +\xi
x_{1}x_{2},
\end{equation}
clearly exhibits permutation symmetry in addition to parity invariance
(exactly like the Hamiltonian of a two-electron system). The advantage of
this model is that the Schr\"{o}dinger equation is exactly solvable. In
fact, by means of the change of variables
\begin{eqnarray}
x_{1} &=&\frac{1}{\sqrt{2}}\left( y_{1}+y_{2}\right) ,\;x_{2}=\frac{1}{\sqrt{%
2}}\left( y_{2}-y_{1}\right) ,  \nonumber \\
y_{1} &=&\frac{1}{\sqrt{2}}\left( x_{1}-x_{2}\right) ,\;y_{2}=\frac{1}{\sqrt{%
2}}\left( x_{1}+x_{2}\right) ,
\end{eqnarray}
the Hamiltonian becomes
\begin{equation}
H=-\frac{1}{2}\left( \frac{\partial ^{2}}{y_{1}^{2}}+\frac{\partial ^{2}}{%
y_{2}^{2}}\right) +\frac{1}{2}\left[ (1-\xi )y_{1}^{2}+(1+\xi
)y_{2}^{2}\right] .
\end{equation}
We appreciate that there are bound states when $-1<\xi <1$. The
corresponding eigenfunctions and eigenvalues are given by
\begin{eqnarray}
\psi _{n_{1}n_{2}}(x_{1},x_{2}) &=&N_{n_{1}n_{2}}\mathit{H}_{n_{1}}\left(
\left[ 1-\xi \right] ^{1/4}y_{1}\right) \mathit{H}_{n_{2}}\left( \left[
1+\xi \right] ^{1/4}y_{2}\right) \times   \nonumber \\
&&\exp \left[ -\frac{\sqrt{(1-\xi )}}{2}y_{1}^{2}-\frac{\sqrt{(1+\xi )}}{2}%
y_{2}^{2}\right] ,  \nonumber \\
E_{n_{1}n_{2}} &=&\sqrt{(1-\xi )}\left( n_{1}+\frac{1}{2}\right) +\sqrt{%
(1+\xi )}\left( n_{2}+\frac{1}{2}\right) ,\;n_{1},n_{2}=0,1,\ldots ,
\nonumber \\
&&
\end{eqnarray}
where $N_{n_{1}n_{2}}$ is a suitable normalization factor that is not
relevant for present purposes and $\mathit{H}_{n}(q)$ is a Hermite
polynomial. Since $P_{12}y_{1}=-y_{1}$ and $P_{12}y_{2}=y_{2}$ we conclude
that every eigenfunction is either symmetric or antisymmetric: $P_{12}\psi
_{n_{1}n_{2}}(x_{1},x_{2})=(-1)^{n_{1}}\psi _{n_{1}n_{2}}(x_{1},x_{2})$.
Also note that $P\psi _{n_{1}n_{2}}(x_{1},x_{2})=\psi
_{n_{1}n_{2}}(-x_{1},-x_{2})=(-1)^{n_{1}+n_{2}}\psi
_{n_{1}n_{2}}(x_{1},x_{2})$ as argued in the preceding section.

In this case it is possible to obtain antisymmetric wavefunctions
of the form $\psi _{space}\psi _{spin}$ for all the eigenfunctions
$\psi _{n_{1}n_{2}}$ of the non-relativistic Hamiltonian. If $\psi
_{space}$ is symmetric, then $\psi _{spin}$ is antisymmetric and
one obtains a singlet state. If, on the other hand, $\psi
_{space}$ is antisymmetric, then we have three symmetric spin
functions and the resulting products give rise to a triplet.
Nascimento\cite{N19} chose only two-electron examples to
illustrate his proposal and therefore his conclusions appeared to
be sound.

\subsection{$N=3$ toy model}

The Hamiltonian
\begin{equation}
H=-\frac{1}{2}\left( \frac{\partial ^{2}}{\partial x_{1}^{2}}+\frac{\partial
^{2}}{\partial x_{2}^{2}}+\frac{\partial ^{2}}{\partial x_{3}^{2}}\right) +%
\frac{1}{2}\left( x_{1}^{2}+x_{2}^{2}+x_{3}^{2}\right) +\xi \left(
x_{1}x_{2}+x_{1}x_{3}+x_{2}x_{3}\right) ,
\end{equation}
exhibits $S_{3}$ permutation symmetry and is parity invariant. It can be
exactly solved by means of the change of variables
\begin{eqnarray}
x_{1} &=&\frac{\sqrt{6}y_{2}}{3}+\frac{\sqrt{3}y_{3}}{3},\;x_{2}=\frac{\sqrt{%
2}y_{1}}{2}-\frac{\sqrt{6}y_{2}}{6}+\frac{\sqrt{3}y_{3}}{3},\;x_{3}=-\frac{%
\sqrt{2}y_{1}}{2}-\frac{\sqrt{6}y_{2}}{6}+\frac{\sqrt{3}y_{3}}{3},  \nonumber
\\
y_{1} &=&\frac{\sqrt{2}x_{2}}{2}-\frac{\sqrt{2}x_{3}}{2},\;y_{2}=\frac{\sqrt{%
6}\left( 2x_{1}-x_{2}-x_{3}\right) }{6},\;y_{3}=\frac{\sqrt{3}\left(
x_{1}+x_{2}+x_{3}\right) }{3},
\end{eqnarray}
that leads to
\begin{equation}
H=-\frac{1}{2}\left( \frac{\partial ^{2}}{\partial y_{1}^{2}}+\frac{\partial
^{2}}{\partial y_{2}^{2}}+\frac{\partial ^{2}}{\partial y_{3}^{2}}\right) +%
\frac{1-\xi }{2}\left( y_{1}^{2}+y_{2}^{2}\right) +\frac{1+2\xi }{2}%
y_{3}^{2}.
\end{equation}
We appreciate that there are bound states provided that $-1/2<\xi <1$. Under
this condition the eigenfunctions and eigenvalues are given by
\begin{eqnarray}
\psi _{n_{1}n_{2}n_{3}}(x_{1},x_{2},x_{3}) &=&N_{n_{1}n_{2}n_{3}}\mathit{H}%
_{n_{1}}\left( \left[ 1-\xi \right] ^{1/4}y_{1}\right) \mathit{H}%
_{n_{2}}\left( \left[ 1-\xi \right] ^{1/4}y_{2}\right) \times  \nonumber \\
&&\mathit{H}_{n_{3}}\left( \left[ 1+2\xi \right] ^{1/4}y_{3}\right) \times
\nonumber \\
&&\exp \left[ -\frac{\sqrt{(1-\xi )}}{2}\left( y_{1}^{2}+y_{2}^{2}\right) -%
\frac{\sqrt{(1+2\xi )}}{2}y_{3}^{2}\right] .  \nonumber \\
E_{n_{1}n_{2}n_{3}} &=&\sqrt{(1-\xi )}\left( n_{1}+n_{2}+1\right) +\sqrt{%
(1+2\xi )}\left( n_{3}+\frac{1}{2}\right) ,\;  \nonumber \\
&&n_{1},n_{2},n_{3}=0,1,2,\ldots .
\end{eqnarray}

Since $y_{1}^{2}+y_{2}^{2}$ and $y_{3}^{2}$ are invariant under permutation,
then the symmetry of an eigenfunction is determined by the product of the
three Hermite polynomials. Thus, the ground state is symmetric as expected.
The excited non-degenerate state $\psi _{001}$ is also symmetric. On the
other hand, the 2-fold degenerate states $\psi _{100}$ and $\psi _{010}$ are
neither symmetric nor antisymmetric and it is not difficult to convince
oneself that there are no linear combinations of $y_{1}$ and $y_{2}$ that
are simultaneous eigenfunctions of the three transposition operators $P_{ij}$%
. The reason is that these variables are basis functions for the irreducible
representation $E$ while the symmetric and antisymmetric functions are basis
for $A_{1}$ and $A_{2}$, respectively, as argued above. Therefore, $%
P_{A_{i}}\psi _{100}=0$ and $P_{A_{i}}\psi _{010}=0$.

Since $Py_{j}=-y_{j}$ then $P\psi _{n_{1}n_{2}n_{3}}(x_{1},x_{2},x_{3})=\psi
_{n_{1}n_{2}n_{3}}(-x_{1},-x_{2},-x_{3})=(-1)^{n_{1}+n_{2}+n_{3}}\psi
_{n_{1}n_{2}n_{3}}(x_{1},x_{2},x_{3})$ in agreement with the result of
section~\ref{sec:Perm sym}.

In this case it is not possible to construct antisymmetric functions of the
form $\psi _{space}\psi _{spin}$ for all the states. First, if $\psi _{space}
$ is basis for the irreducible representation $E$ it is neither symmetric
nor antisymmetric. Second, if $\psi _{space}$ is basis for $A_{1}$ we cannot
construct a three-electron antisymmetric spin function (see below). The only
case in which we can obtain antisymmetric functions of the form $\psi
_{space}\psi _{spin}$ is when $\psi _{space}$ is basis for $A_{2}$ and $\psi
_{spin}=\omega (s_{1})\omega (s_{2})\omega (s_{3})$ with $\omega $ being
either $\alpha $ or $\beta $ (the one-electron spin functions for $m_{s}=1/2$
or $m_{s}=-1/2$, respectively).

According to the principles of quantum mechanics the states of the
system should be basis for $A_{2}$. We can obtain such states by
forcing the antisymmetry on products of space times spin
functions; for example by means of the expression $P_{A_{2}}\psi
_{n_{1}n_{2}n_{3}}(x_{1},x_{2},x_{3})\omega _{i}(s_{1})\omega
_{j}(s_{2})\omega _{k}(s_{3})$, where $P_{A_{2}}$ acts upon both
spatial and spin labels. It seems paradoxical that if $\psi
_{n_{1}n_{2}n_{3}}(x_{1},x_{2},x_{3})$ is basis for $A_{1}$ then $%
P_{A_{2}}\psi _{n_{1}n_{2}n_{3}}(x_{1},x_{2},x_{3})\omega _{i}(s_{1})\omega
_{j}(s_{2})\omega _{k}(s_{3})=\psi
_{n_{1}n_{2}n_{3}}(x_{1},x_{2},x_{3})P_{A_{2}}\omega _{i}(s_{1})\omega
_{j}(s_{2})\omega _{k}(s_{3})=0$. The symmetric spatial functions, one of
those proposed by Nascimento\cite{N19}, are not allowed by the principles of
quantum mechanics. Furthermore, a configuration interaction calculation
based on a linear combination of suitable Slater determinants will not give
the energy levels $E_{00n}$ because the corresponding eigenfunctions $\psi
_{00n}$ are basis for the irreducible representation $A_{1}$.

It is worth noting that the conclusion just drawn applies to any realistic
three-dimensional $N$-electron model because the argument is based only on
the permutation of the variables in the wavefunction and is, therefore,
model independent (the irreducible representation $A_{1}$ appears in any
symmetric group $S_{N}$).

\section{Further comments and conclusions}

\label{sec:conclusions}

In the preceding two sections we have clearly shown that the eigenfunctions $%
\psi $ of a non-relativistic Hamiltonian $H$ for a system of $N>2$ identical
particles do not satisfy $P_{ij}\psi =\pm \psi $ for all $i$,$j$ and all $%
\psi $. There are always two one-dimensional irreducible representations of $%
S_{N}$ that satisfy this requirement but the basis for the
remaining representations are neither symmetric nor antisymmetric.
For this reason it is not possible to construct antisymmetric
functions of the form $\psi _{space}\psi _{spin}$ except for the
states with $\left| M_{S}\right| =\max \{S\}$.
Nascimento\cite{N19} considered only two-electron examples where
such factorization is possible for all states because there is
only one transposition operator and $S_{2}$ exhibits only two
one-dimensional irreducible representations, the basis of which
are symmetric and antisymmetric functions.

Permutation symmetry is a well-defined mathematical concept. In the case of $%
N=2$ particles the operators $S=\frac{1}{2}\left( 1+P_{12}\right) $ and $A=%
\frac{1}{2}\left( 1-P_{12}\right) $ project onto the spaces of symmetric and
antisymmetric functions, respectively\cite{CTDL77}. For example: $S\phi
_{1}(1)\phi _{2}(2)=\frac{1}{2}\left[ \phi _{1}(1)\phi _{2}(2)+\phi
_{2}(1)\phi _{1}(2)\right] $, $A\phi _{1}(1)\phi _{2}(2)=\frac{1}{2}\left[
\phi _{1}(1)\phi _{2}(2)-\phi _{2}(1)\phi _{1}(2)\right] $, $S\phi (1)\phi
(2)=\phi (1)\phi (2)$ and $A\phi (1)\phi (2)=0$. Therefore, from a strictly
mathematical point of view $\phi (1)\phi (2)$ is a well-defined symmetric
function\cite{CTDL77}. Curiously, Nascimento\cite{N19} considers the
permutation operator to be ill-defined in the latter case, a statement that
obviously has no mathematical support.

In the case of helium-like atoms we may obtain a reasonable approximation to
the ground state by means of an approximate trial function of the form $%
\varphi (\eta )=1S(\eta r_{1})1S(\eta r_{2})$, where $1S$ is a hydrogen-like
atomic orbital and $\eta $ an effective nuclear charge to be optimized by
means of the variational method\cite{P68}. Eckart\cite{E30} proposed the
trial function $\varphi (\alpha ,\beta )=1S(\alpha r_{1})1S(\beta
r_{2})+1S(\beta r_{1})1S(\alpha r_{2})$ that is obviously better because it
has two adjustable parameters and reduces to the previous one when $\alpha
=\beta $. Consequently, it is not surprising that the latter yields a lower,
and therefore better, energy than the former. From a strict mathematical
point of view both functions are symmetric because $P_{12}\varphi =\varphi $%
. In this case, variational flexibility and not lack of permutation symmetry
(as argued by Nascimento) is the cause of the different performances.

We can carry out a similar analysis of the comparison of molecular orbital
and valence bond methods in the case of the ground-state of the hydrogen
molecule in the Born-Oppenheimer approximation. Both approximate space
functions are symmetric under permutation of the electrons but the valence
bond function was constructed to yield the exact result in the dissociation
limit\cite{P68}. It is therefore no surprising its striking better
performance at large internuclear distances. Once again, the lack of
permutation symmetry invoked by Nascimento does not appear to be the issue.

\end{document}